\documentclass[10pt,twocolumn,aps,pra,longbibliography]{revtex4-2}
\usepackage[utf8]{inputenc}

\usepackage[T1]{fontenc}
\usepackage{xcolor}
\usepackage{braket}
\usepackage{soul}
\usepackage{circuitikz}
\usepackage{amsmath}
\usepackage{amssymb}
\usepackage{graphicx}
\usepackage[english]{babel}
\usepackage{float}

\newcommand{\iu}{\mathrm{i}}

\begin{document}

\title{Topologically Protected  Edge States in a Single Flux Qubit}

\author{P. Pérez-Fernández}
\affiliation{Universidad de Salamanca, E-37008 Salamanca, Spain}

\author{Manuel Pino}
\affiliation{Universidad de Salamanca IUFFyM, E-37008 Salamanca, Spain}

\date{\today}

\begin{abstract}
We show that topological edge states can be realized in a simple superconducting circuit composed of an inductively shunted flux qubit. The Josephson potential of that qubit forms a honeycomb lattice similar to the case of graphene. The additional shunting element produces a quartic potential along one flux direction, which effectively creates a boundary in flux space. We map analytically the low-energy dynamics of the circuit to an extended Su--Schrieffer--Heeger (SSH) chain, in which a quartic potential is added to the usual SSH Hamiltonian. This mapping is confirmed by numerical simulations of the full superconducting circuit. We argue that chiral symmetry still holds for the edge states of the extended model around a chiral point in parameter space. Close to this point, we numerically find that edge states in our superconducting circuit are topologically protected against perturbations that obey chiral symmetry, as is the case for charge noise. Our results show that topological effects previously thought to occur in multi-qubit systems, such as protection against decoherence, can also be found in much simpler single superconducting qubits.
\end{abstract}

\maketitle

\emph{Introduction.--} Topological properties are crucial to understand a broad family of materials, most famously exemplified by the quantum Hall effect \cite{klitzing_new_1980}. Those properties are distinguished by global, integer-valued invariants of their band structure. These are quantities, such as the Chern number or the
winding number, that remain unchanged under continuous deformations of the Hamiltonian that respect a given set of symmetries~\cite{hasan_colloquium_2010}. A direct consequence of this rigidity, applied to a system described by a Hamiltonian with non-trivial topology, is the bulk-boundary correspondence.
Topological boundary modes appear whenever open boundary conditions are imposed on such a system~\cite{hatsugai_chern_1993, essin_bulk-boundary_2011}. Besides their intrinsic interest, protected topological edge modes can be used to build a fault-tolerant quantum computer~\cite{kitaev_unpaired_2001, pino_unpaired_2015, zanardi_topological_2003, estarellas_topologically_2017}.

Because of their relevance for quantum computing, there have been many attempts to build topological edge modes in several quantum hardware platforms, such as superconducting circuits~\cite{ripoll_quantum_2022}.
For instance, the SSH and the Harper--Hofstadter models of the integer quantum Hall effect have been implemented on chains of coupled superconducting qubits~\cite{xue_topological_2021}.
The resulting edge states have in turn been shown to be useful for quantum information processing, such as quantum state transfer~\cite{mei_robust_2018}, encoding a protected computational basis~\cite{hu_realizing_2017}, and realizing non-Abelian gate operations~\cite{cao_detecting_2020}.
In all these cases, however, a chain of several coupled qubits or resonators is required to host the \emph{real-space} edge modes. This adds a significant layer of complexity to the design, control, manipulation and coupling of the qubits. A natural question is then whether a simpler superconducting circuit can host edge modes without the burden of coupling several qubits together.

\begin{figure*}[t]
    \centering
    \begin{minipage}[b]{0.48\textwidth}
        \raggedright
        \textbf{a)}\par\vspace{2pt}
        \centering
        \begin{tikzpicture}[scale=1, transform shape]
        \draw (2.2,2.9) to[openbarrier, name=Ja] (4.6,2.9);
        \draw (2.2,2.9) to[openbarrier, name=Jl] (2.2,0);
        \draw (4.6,2.9) to[openbarrier, name=Jr] (4.6,0);
        \draw (2.2,0) -- (4.6,0);
        \node[ground] at (3.4,0) {};
        \def\jjh{0.147}
        \foreach \J in {Ja,Jl,Jr}
            \draw ($(\J.center)+(-\jjh,-\jjh)$) rectangle ($(\J.center)+(\jjh,\jjh)$);
        \node[below=2pt] at (Ja.center) {$\alpha E_J$};
        \node[right=2pt] at (Jl.center) {$E_J$};
        \node[right=2pt] at (Jr.center) {$E_J$};
        \node at (3.4,1.45) {$\Phi$};
        \draw[->,thick] (3.4,1.45) ++(-50:0.4) arc (-50:255:0.4);
        \draw (2.2,2.9) to[capacitor, l_=$C_g$] (2.2,3.9) node[above] {$V_{g1}$};
        \draw (4.6,2.9) to[capacitor, l_=$C_g$] (4.6,3.9) node[above] {$V_{g2}$};
        \node[circ] at (2.2,2.9) {};   \node[circ] at (4.6,2.9) {};
        \node[below left=1pt]  at (2.2,2.9) {$1$};
        \node[below right=1pt] at (4.6,2.9) {$2$};
        \draw (2.2,2.9) -- (0.3,2.9);
        \draw (2.2,0)   -- (0.3,0);
        \draw (0.3,2.9) -- (0.3,0);   \draw (1.5,2.9) -- (1.5,0);
        \begin{scope}[circuitikz/bipoles/openbarrier/gap=0]
          \foreach \y in {2.65,1.85,1.05}
              \draw (0.3,\y) to[openbarrier] (0.3,\y-0.8);
          \foreach \y in {2.65,2.25,1.85,1.45,1.05,0.65}
              \draw (1.5,\y) to[openbarrier] (1.5,\y-0.4);
        \end{scope}
        \node[above] at (0.3,2.9) {$n$};
        \node[above] at (1.5,2.9) {$\Omega n$};
        \node at (0.9,1.45) {$\Phi'$};
        \draw[->,thick] (0.9,1.45) ++(-50:0.33) arc (-50:255:0.33);
        \end{tikzpicture}
    \end{minipage}
    \hfill
    \begin{minipage}[b]{0.48\textwidth}
        \raggedright
        \textbf{b)}\par\vspace{2pt}
        \centering
        \includegraphics[width=0.8\linewidth]{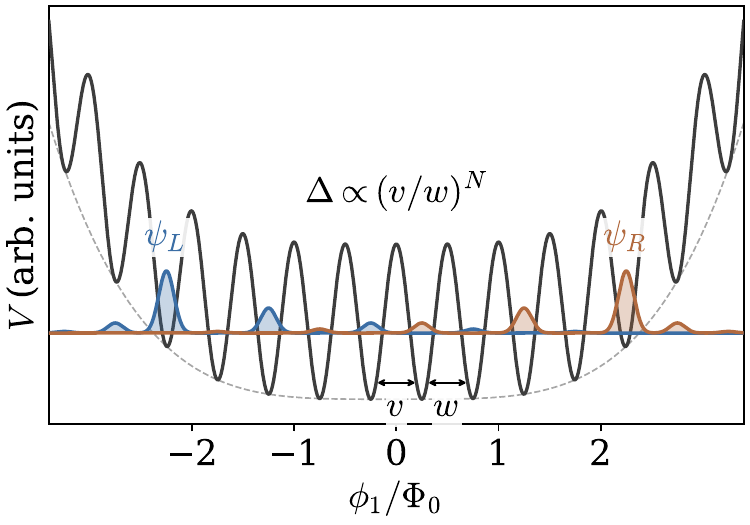}
    \end{minipage}
   \caption{\textbf{a)} A three-Josephson-junction flux qubit, composed of a superconducting loop with two junctions of energy $E_J$ and one of energy $\alpha E_J$, threaded by an external flux $\Phi$. Nodes $1$ and $2$ are biased by the gate voltages $V_{g1}$ and $V_{g2}$ through the capacitances $C_g$. Node $1$ is shunted by two parallel arrays of $n$ and $\Omega n$ junctions of energy $E_J'$ enclosing a flux $\Phi'$, which contribute a quartic potential in $\phi_1$. The loop between this element and the qubit is threaded by no flux. \textbf{b)} Sketch of the phenomenology expected for the edge states in the superconducting circuit. The washboard potential of the 3JJ loop is confined by the quartic potential (dashed grey) along the $\phi_1$ direction. For $|w|>|v|$, states localized at the boundary emerge within the gap. The level splitting and the decay length of the edge state are exaggerated for clarity, and the well spacing is drawn uniform, whereas the actual lattice alternates intracell and intercell separations.}
    \label{fig: circuit_and_ideal}
\end{figure*}

Here, we propose to engineer a topological edge state that lives in a single superconducting qubit circuit. This may seem contradictory at first, since edge states are usually thought of as living at distinct real-space locations. To achieve such an edge state in a single qubit, we work in the effectively infinite two-dimensional phase space of a flux qubit~\cite{orlando_superconducting_1999}. A quartic potential is then added in one of the phase variables by shunting the qubit with two parallel Josephson-junction arrays, see Fig.~\ref{fig: circuit_and_ideal}a, similarly to recent blochnium and fluxonium proposals~\cite{manucharyan_fluxonium_2009, koch_charging_2009, hita-perez_bound_2022, chirolli_quartic_2023}. Specifically, we map the superconducting-circuit Hamiltonian onto a finite-size Su--Schrieffer--Heeger (SSH) model \cite{su_soliton_1980, zak_berrys_1989}, in close analogy with the theory developed for graphene nanoribbons \cite{cao_topological_2017}. We will show that the resulting edge modes closely resemble topological ones and, furthermore, that they display a high degree of resilience against charge noise. Previous works have reported topological properties in similar superconducting circuits~\cite{berry_quantal_1984, leek_observation_2007, fatemi_weyl_2021, peyruchat_spectral_2024}. None of them, however, reports a protected topological edge state in a single qubit, which is our main result.

\emph{Superconducting Circuit Hamiltonian.--} The circuit we propose consists of an inductively shunted three-Josephson-junction (3JJ) qubit. The qubit loop is interrupted by three Josephson junctions. Two of them are identical with Josephson energy $E_J$, while the third has an area $\alpha$ times that of the other two, which will be fixed to $\alpha=1$ \cite{orlando_superconducting_1999, mooij_josephson_1999}. This loop is additionally shunted by two parallel arrays of $n$ and $\Omega n$ Josephson junctions, each junction having energy $E_J'$. From the circuit representation in Fig.~\ref{fig: circuit_and_ideal}a, we can derive the following Hamiltonian \cite{devoret_quantum_1997, ripoll_quantum_2022}
\begin{equation}
    H = \frac{1}{2}(\boldsymbol{q} + \boldsymbol{q}_g)^T\boldsymbol{C}^{-1}(\boldsymbol{q} + \boldsymbol{q}_g)+ U_{3JJ} + U_{sh},
    \label{eq: H_circuit}
\end{equation}
where the capacitance matrix has diagonal elements $\boldsymbol{C}_{ii}= C(1 + \alpha + \gamma)$ and off-diagonal elements $\boldsymbol{C}_{ij}= -C\alpha$, with $\gamma = C_g/C$.
The potentials of the 3JJ and of the shunting element are, respectively,
\begin{align}
  \frac{U_{3JJ}}{E_J} =&
    - \alpha \cos\!\left(\frac{\Phi - \phi_1 + \phi_2}{\varphi_0}\right)
    -\sum_{i=1,2}\cos\!\left(\frac{\phi_i}{\varphi_0}\right),\label{eq: U_3JJ}\\
    \frac{U_{sh}}{E'_J} = &\left[\frac{n}{\Omega}\cos\left(\frac{\phi_1}{n\varphi_0}\right) - n\Omega\cos\left(\frac{\phi_1}{n\Omega\varphi_0}\right)\right],\label{eq: U_sh}
\end{align}
where $\varphi_0=\Phi_0/2\pi$, with $\Phi_0$ the flux quantum. The loop formed by the two arrays is threaded by the flux $\Phi'=\Phi_0/2$, and $\Phi$ is the flux threading the 3JJ qubit. The symbols $\phi_1$ and $\phi_2$ are, respectively, the flux variables at nodes 1 and 2, with canonically conjugate charges $\boldsymbol{q}=(q_1, q_2)^{T}$ that fulfill $[\phi_i,q_j]=\iu\hbar\,\delta_{ij}$. The offset charges are defined as $\boldsymbol{q}_g=(q_{g1},q_{g2})^{T}$ and $q_{gi}=\gamma C V_{gi},$ where $V_{gi}$ is the gate voltage applied to the node $i.$ The shunting element is composed of two parallel branches, one of which has $\Omega$ times more junctions than the other (see Fig.~\ref{fig: circuit_and_ideal}a). They create an effective quartic potential $U_{sh}$, as in a quartic superinductor~\cite{chirolli_quartic_2023}, once their internal degrees of freedom are disregarded within the Born--Oppenheimer approximation \cite{randeria_dephasing_2024, ardati_using_2024}.

\emph{Effective SSH model.--} Having introduced the circuit, we now derive its effective low-energy Hamiltonian, in close analogy with graphene nanoribbons \cite{cao_topological_2017}.
For $\alpha=1$ and $\Phi=\Phi_0/2$ the 3JJ qubit potential $U_{3JJ}$ in Eq.~\eqref{eq: U_3JJ} is periodic in both flux variables with period $2\pi\varphi_0.$ Its minima form a two-dimensional honeycomb lattice with two sublattices, denoted by $L$ and $R$, which are the periodic repetitions of $L\equiv(-\pi/3,+\pi/3)\varphi_0$ and $R\equiv(+\pi/3,-\pi/3)\varphi_0$~\cite{hita2021three}. When $E_C/E_J\ll 1,$ the low-energy dynamics reduces to tunneling between nearest-neighbor wells of this lattice. At our working point $\alpha=1,$ the amplitudes of the two tunneling paths are almost identical, and we describe both by a single amplitude $t$. Due to the periodicity in the flux variables \cite{koch_charge-insensitive_2007}, the dimensionless offset charges $k_i\equiv -q_{g,i}/2e$, conjugate to $\phi_1/\varphi_0$ and $\phi_2/\varphi_0$, play the role of quasimomenta, with $k_1, k_2 \in [-1/2, 1/2)$ \cite{ashcroft_solid_1976}.
The resulting band structure is that of graphene \cite{wallace_band_1947}, with two Dirac cones located at $(k_1,k_2)=\pm(1/3,\,-1/3)$~\cite{chirolli_full_2006, tiwari_suppression_2007}. The shunting potential $U_{sh}(\phi_1)$ in Eq.~\eqref{eq: H_circuit} effectively truncates the graphene lattice to a finite strip, as in a graphene nanoribbon \cite{cao_topological_2017}. Periodicity along $\phi_2$ is preserved, so $k_2$ remains a good quantum number, whereas translational invariance along $\phi_1$ is effectively lost, similarly to the case of the fluxonium \cite{note_tunneling}.

We derive an effective tight-binding Hamiltonian along the $\phi_1$ direction, with each cell holding one $R$ and one $L$ site, by Fourier transforming along the periodic direction $\phi_2.$ Doing so, we obtain a one-dimensional SSH Hamiltonian composed of the intracell and intercell tunnelings $H_v=-v(k_2)\,\sum_{i}\hat{l}^\dagger_{k_2,i}\hat{r}_{k_2,i}+\text{h.c.}$ and
$H_w=-w\,\sum_{i}\hat{r}^\dagger_{k_2,i+1}\hat{l}_{k_2,i}+\text{h.c.}$, plus the quartic potential. The operators $\hat{r}^{\dagger}_{k_2,i}$ and $\hat{l}^{\dagger}_{k_2,i}$ create Bloch states with momentum $k_2$ along the $\phi_2$ direction but localized in the $\phi_1$ direction at the right and left minimum of the $i$-th unit cell, respectively. The effective Hamiltonian reads
\begin{equation}
    H_{\rm eff}(k_2)=H_w + H_v(k_2)+V,\label{eq:tight binding hamiltonian for ribbon}
\end{equation}
where $V=\sum_{i} \epsilon_{R_i}\hat{r}^\dagger_{k_2,i}\hat{r}_{k_2,i}+ \epsilon_{L_i}\hat{l}^\dagger_{k_2,i}\hat{l}_{k_2,i}$, with energies $\epsilon_{R_i}=\lambda V_4(i-\tfrac 16)$ and $\epsilon_{L_i}=\lambda V_4(i+\tfrac16)$ given by the quartic potential $V_4(u)=(2\pi u)^4+\mathcal{O}(u^6)$. The hoppings are $v(k_2)=t\left(1+e^{\iu 2\pi k_2}\right)$ and $w=t$, with $t=-\bra{R_{i}}H(E_J'=0)\ket{L_{i}}$, where $\ket{L_{i}}$ and $\ket{R_{i}}$ are the states localized at the corresponding minima of a single unit cell and $H(E_J'=0)$ is the Hamiltonian of the circuit without the shunting element. For the hopping amplitude, the WKB approximation gives $t\propto e^{-\sqrt{8E_J/E_C}}$. For the confinement strength, expanding Eq.~\eqref{eq: U_sh} around $\phi_1=0$ we obtain $\lambda \approx E'_J/(24\,n^{3})\left(1/\Omega-1/\Omega^{3}\right)$, while the quadratic term cancels identically for any $\Omega$, which is precisely the purpose of the two-array construction.

We have not yet settled the length of the effective SSH chain in Eq.~\eqref{eq:tight binding hamiltonian for ribbon}. We define it so that the edge states of the effective chain are the in-gap states of the circuit that most closely resemble the edge states of the SSH model, which, as shown below, are those closest to the chiral point. The effective chain is then formed by the $N+1$ cells $i=-N/2,\dots,N/2$, with $N$ even, whose two ends host those states.

\emph{Topological Edge States.--} Equation~\eqref{eq:tight binding hamiltonian for ribbon} is the Hamiltonian of the SSH model \cite{su_soliton_1980}, with intracell tunneling $H_v$, controlled by $v(k_2)$, and intercell tunneling $H_w$, whose amplitude is $w$, plus a potential $V$. If that potential were zero, edge states would appear when $|v|<|w|$, which here corresponds to $1/3<|k_2|\leq1/2$ in our model. The symmetry protecting those topological states in the pure SSH model is the chiral symmetry
\begin{equation}
    \Gamma\,(H_v+H_w)\,\Gamma=-(H_v+H_w),\qquad \Gamma=1-2P_R,
    \label{eq:chiral}
\end{equation}
with $P_R$ the projector onto the $R$ sublattice, but the on-site term satisfies $\Gamma V\Gamma=V$ and breaks this symmetry \cite{asboth_short_2016}. Neither the degeneracy of the edge states nor their confinement to a single sublattice is then exact, although both are hallmarks of the topology of the system. We now show that both survive in our full effective model, Eq.~\eqref{eq:tight binding hamiltonian for ribbon}, near $k_2=\pm1/2$ with very good accuracy.

At $k_2=\pm1/2$ one has $v=0$, and the chain splits exactly into dimers $(L_i,R_{i+1})$ coupled by $w$. Where the confinement is flat, the dimers are symmetric and form the bulk bands of the SSH chain. Where it rises, the detuning between the two sites of a dimer exceeds $w$, and its lower state $\ket{-,i}$ lies almost entirely on $L_i$, a single site of a single sublattice. The confinement thus does not cut the bond at the end of the chain but renders it ineffective by detuning, and the dimer $N/2$ terminates the effective chain. Together with its mirror partner at the opposite end, it forms the pair of edge states.

\begin{figure}[t]
    \centering
        \includegraphics[width=1.\linewidth]{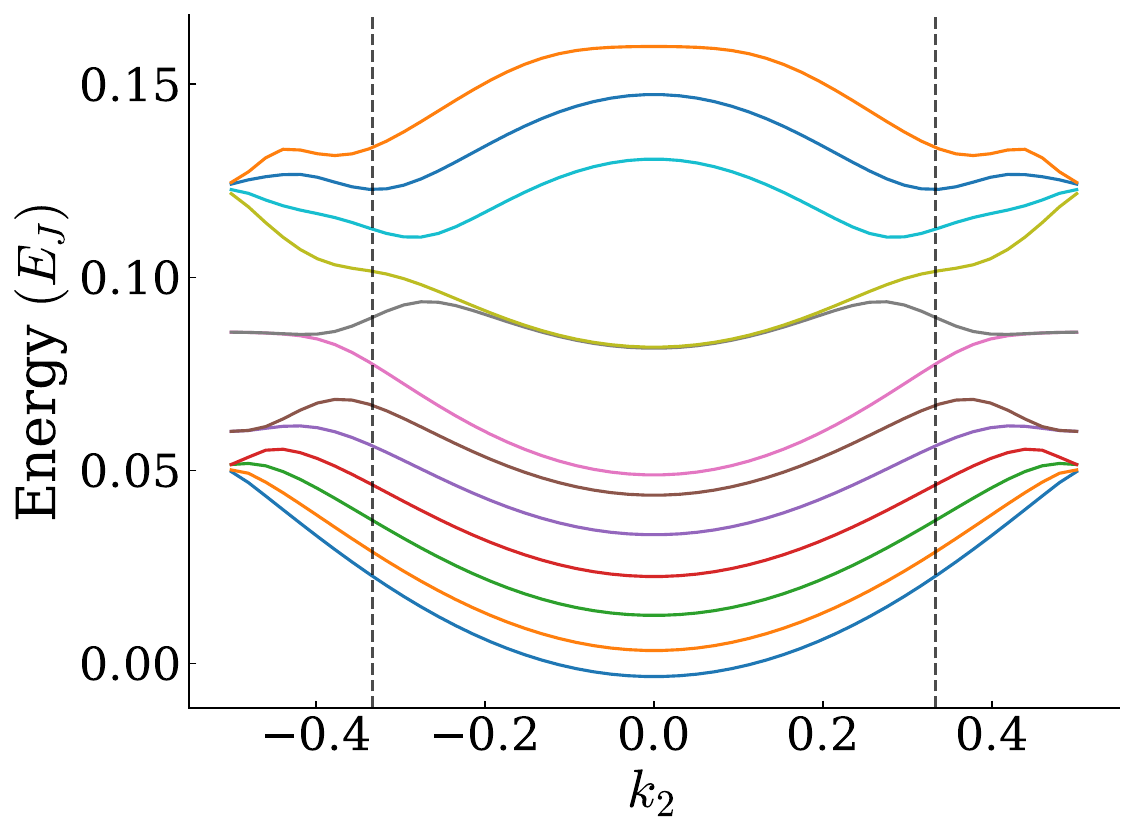}
    \caption{Numerically obtained low-energy spectrum as a function of the quasimomentum $k_2$ for the superconducting circuit composed of a 3JJ qubit and the shunting element. The circuit parameters are $E_J/E_C=30$, $E_J'=10\,E_J$, $n=100$, and $\Omega=2$.
    The two vertical lines mark $k_2=\pm1/3$, where the topological transition is expected.
    The levels begin to split somewhat before $|k_2|=1/3$ owing to finite-size hybridization of the two edge states.
    }
    \label{fig: circuit results}
\end{figure}

Treating $H_v$ as a perturbation of these dimer states, the edge state $\ket{\psi_{N/2}}$ spreads onto the neighboring sites $L_{N/2-1}$ and $R_{N/2}$. To leading order, its energy and its weight on the wrong sublattice within the effective chain are
\begin{align}
    &E(k_2)=E+|v|^{2}\cos^{2}\varphi_{N/2}\,\frac{S}{D}+\mathcal{O}(v^4),\label{eq:Ek2}\\
    &\bra{\psi_{N/2}}P_R\ket{\psi_{N/2}}_{\rm int}=\frac{|v|^{2}S^{2}/D^{2}}{1+|v|^{2}\left(w^{2}+S^{2}\right)/D^{2}},\label{eq:PRint}
\end{align}
where $|v|^2=4t^2\cos^2(\pi k_2)$, $E$ is the unperturbed energy of the pair, $\cos^2\varphi_{N/2}$ its weight on $L_{N/2}$, and $D$ the product of the two energy denominators of the neighboring dimer. The subscript ${\rm int}$ indicates that the projection is restricted to the $R$ sites inside the effective chain, $R_i$ with $i\le N/2$, with the state normalized there. The site $R_{N/2+1}$ is excluded because it lies beyond the chain, where the state keeps a weight $\sin^2\varphi_{N/2}$ that is a boundary effect of the quartic potential rather than part of the SSH problem. Both the flatness of the band and the purity of the sublattice are therefore controlled by the single parameter
\begin{equation}
    S=E-\epsilon_{L_{N/2-1}},\label{eq:S}
\end{equation}
the detuning between the edge level and the nearest site of its own sublattice. When $S=0$ the pair is flat and sublattice-polarized up to corrections of order $v^4$, and behaves as a genuine SSH edge state. We refer to $S=0$ as the chiral point of the circuit.

\emph{Numerical Simulation of the Edge States.--} We now turn to the numerical simulation of the superconducting circuit given by the Hamiltonian in Eq.~\eqref{eq: H_circuit}. We aim to have a circuit such that the chiral condition $S=0$, for $S$ defined in Eq.~\eqref{eq:S}, is met for the states within the gap between the two bands of the effective SSH model. We have specifically chosen the circuit parameters so that the effective chain length is $N=6$, with experimentally sensible junction parameters in the shunting element, $E'_J=10\,E_J$, $n=100$ and $\Omega=2$ (where $E_J$ is the 3JJ qubit junction energy). This choice produces a confinement strength in Eq.~\eqref{eq: U_sh} of $\lambda=E'_J(1/\Omega-1/\Omega^3)/24n^3=1.56\times10^{-7}E_J.$ With these parameters the chiral condition $S=0$ fixes a chiral hopping $t^\star=0.03\,E_J$, which turns out to produce approximately the ratio $E_J/E_C\approx 30$\footnote{Notice that this last approximation is obtained by using the relation between the hopping and the circuit parameters $t=t (E_J/E_C)$ when there is no shunting element, $E_{J}^\prime=0,$ as obtaining this relation for the full circuit is numerically much harder. We highlight that in our numerical simulation the chiral point is not exactly met. Nevertheless, the topological properties of the system are very robust.}.

Figure~\ref{fig: circuit results} shows the low-energy spectrum as a function of the offset charge $k_2=-\gamma C V_{g2}/(2e)$. Two bands are separated by a gap that hosts the pair of states $6$ and $7$, the flattest in the spectrum. Fitting the spectrum with the tight-binding chain of Eq.~\eqref{eq:tight binding hamiltonian for ribbon} gives $t=0.035\,E_J$ (data not shown), close to the estimate $t^\star=0.03\,E_J$. The wavefunction of the pair (data not shown) is concentrated at the two ends of the effective chain and, at each end, occupies a single sublattice to a very good approximation. As expected, the two states split as $|k_2|$ approaches $1/3$ and the system leaves the topological regime. Since the condition $S=0$ fixes $E_J/E_C$ for each effective length $N$, a chiral point exists for every pair except the lowest one. For our parameters only the pair $6,7$ lies close enough to its chiral point to show robust topological signatures.

\emph{Topological protection.--} We now analyze how the edge doublet, whose two localized states we denote $\ket{0}_q$ and $\ket{1}_q$, is protected against noise. Topological states are protected only against perturbations that preserve the relevant symmetry~\cite{kitaev_unpaired_2001, pino_unpaired_2015, mcginley2017}, here the chiral symmetry of Eq.~\eqref{eq:chiral}. Charge noise on node 2 shifts $k_2$ and therefore modulates only the hopping $v$, which preserves it. Flux noise instead adds an on-site energy of opposite sign on the two sublattices, which breaks it. We therefore expect protection against charge noise but not against flux noise.

\begin{figure}[t!]
    \centering
    \includegraphics[width=\linewidth]{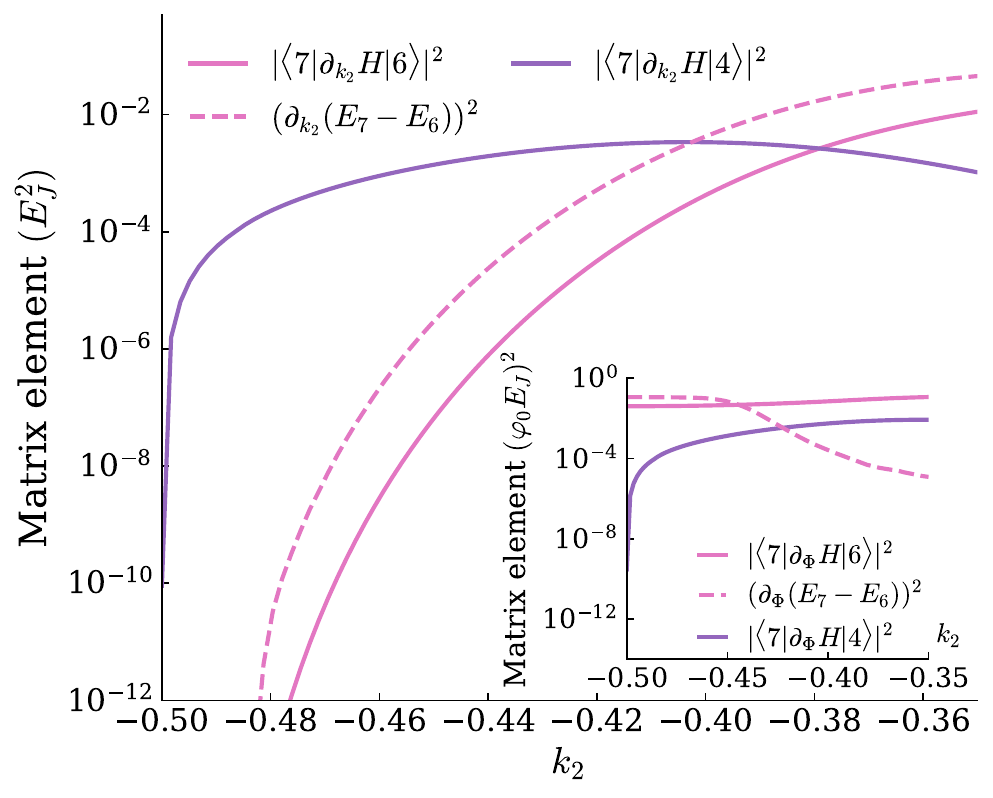}
    \caption{
     Pure dephasing and decay charge matrix elements as a function of $k_2$, in units of $E_J^2$, at $\Phi=0.5\,\Phi_0$.
     The dephasing element is the numerical derivative of the energy splitting with respect to $k_2$.
    Inset: The same for flux noise. The dephasing element is evaluated at $\Phi=(0.5+10^{-5})\Phi_0$, since the derivative vanishes at $\Phi=0.5\,\Phi_0$ by symmetry. The decay element is evaluated in the basis $\frac{1}{\sqrt{2}}(\ket{0}_q\pm\ket{1}_q)$, since $\ket{0}_q$ and $\ket{1}_q$ are, to a good approximation, flux eigenstates.}
    \label{fig:noise}
\end{figure}

We first discuss charge noise. Let us remark that, thanks to the Josephson junction arrays, the qubit is insensitive to charge noise on node 1, as in the fluxonium~\cite{manucharyan_fluxonium_2009}. On node 2, the two edge states reside at opposite ends of the chain, and parity symmetry makes their on-site energies identical to all orders in $v$. Within the tight-binding model, their splitting arises only from the hybridization through the intermediate dimers, at order $N+1$ in $v$, so that the qubit frequency depends on $k_2$ only through $|v|^{N+1}$. This is consistent with the degeneracy of the pair over a wide range of $k_2$ around $\pm1/2$ in Fig.~\ref{fig: circuit results}. Relaxation is suppressed by the sublattice structure, as in bifluxon qubits~\cite{kalashnikov_bifluxon_2020, ardati_using_2024}. Since the two states have support on opposite sublattices, charge noise can only connect them through the residual overlap of Eq.~\eqref{eq:PRint}, which is very small near the chiral point. We can numerically confirm the negligible importance of the charge fluctuations with the data presented in Fig.~\ref{fig:noise}, which shows both the dephasing element, given by the derivative of the doublet splitting with respect to $k_2$, and the decay element. Our data are consistent with an exponential suppression of the coupling of the topological doublet to charge noise in the topological region $-1/2<k_2<-1/3$, a robust confirmation of the protection against perturbations that preserve chiral symmetry. Flux noise, which breaks this symmetry, behaves very differently. Its dephasing and decay elements, shown in the inset of Fig.~\ref{fig:noise}, are not small, and are much larger than the corresponding charge elements.

\emph{Conclusions.--} We have designed an inductively shunted flux qubit with topologically protected edge modes. The inductive shunting creates a steep enough potential so that the phase dynamics of the Josephson junctions occurs in an effective finite system, as its low-energy dynamics is confined by this potential. Our main result is that there are edge modes that retain a strong topological protection against decoherence induced by charge fluctuations. This is a particularly counter-intuitive result, since the single flux qubit that we are using is extremely sensitive to charge noise without the border induced by the shunting element \cite{orlando_superconducting_1999}. The protection we have obtained is not complete, as the system is certainly not protected against flux noise, which breaks the chiral symmetry. This reduced protection against perturbations that break the symmetry underlying the non-trivial topology is not unique to this case. Indeed, Majorana modes in 1D superconductors are likewise not protected against processes that change the fermion parity of the wire, such as quasiparticle poisoning~\cite{kitaev_unpaired_2001,karzig2017scalable,karzig2021quasiparticle}.

The present construction provides a simple route to design topologically protected superconducting circuits in which the lattice, its boundary and its tunneling amplitudes are all set by circuit parameters and gate voltages rather than by the fabrication of many coupled elements. We believe the ideas presented here will be very helpful in the search for topologically protected qubits, a key ingredient of noise-resilient quantum hardware, in much simpler systems than previously thought.

\begin{acknowledgments}
M.~P. acknowledges support through grant no. PID2024-156340NB-I00 funded by Ministerio de Ciencia, Innovación y Universidades/Agencia Estatal de Investigación (MICIU/AEI/10.13039/501100011033) and the European Regional Development Fund (ERDF). The numerical computations were performed in the facilities of Supercomputación Castilla y León (SCAYLE).
\end{acknowledgments}

\bibliography{Bib_fusionada}

\end{document}